# Evaluating Financial Model Performance: An Empirical Analysis of Some North Sea Investments

Grenville J. Croll, David F. Baker, Ola Lawal
grenville@spreadsheetrisks.com

**Fifty North Sea oil & gas investment transactions were analysed using traditional spreadsheet based financial modelling methods. The purpose of the analysis was to determine if there was a statistically significant relationship between the price paid for an oil & gas asset and the actual or expected financial return over the asset's economically useful life. Several interesting and statistically significant relationships were found which reveal useful information about financial modelling performance, the premia paid to acquire North Sea assets, the contribution oil and gas price uncertainty has on estimates of future financial returns and the median financial return of these North Sea Investments.**

## 1. Introduction

There has been considerable debate in the literature regarding the inadequacies of traditional methods of project valuation (Trigeorgis, 1996 & references). New methods of valuation, including in particular the real options methodology have gradually emerged as a means of potentially bridging the perceived gap between the price paid for an exploration & production asset (Siegel, Smith & Paddock, 1987) and its value as determined by traditional (e.g. spreadsheet based) NPV methods. Other more recent research has focussed on the use of advanced decision analysis techniques as a means of extracting value from the challenging investment environment of the oil and gas industry (MacMillan, 2000), particularly in hostile environments such as the North Sea. Some evidence is available that the use of more advanced decision making techniques is correlated with increased corporate value. There is otherwise little empirical information regarding the systematic effectiveness of differing valuation methodologies, financial modelling and decision-making techniques upon project or corporate performance.

The purpose of the present study was to determine if there was a systematic relationship between the price paid for an (unquoted) share of an investment opportunity in the North Sea and its actual or expected return as modelled in a traditional spreadsheet based NPV analysis. The present study is believed to be unique in that data for a whole portfolio of assets was available. This study was part of a wider investigation into the utility of the real options methodology which was undertaken by one of the authors as part of a Masters thesis in petroleum engineering (Lawal, 2001).

## 2. Transaction Data

Data for fifty transactions in the North Sea Oil & Gas Industry was extracted from public domain sources (Andersen, 2001a & 2001b). Data included the sterling price paid (SPP) at the date of the transaction, the date of the transaction, projected and actual production profiles, projected and actual oil and gas prices, estimated costs, exchange rates, inflation, tariffs, royalty, petroleum revenue and corporate taxation. The fifty transactions were all of those for which complete data was available. There was otherwise no pre-selection or filtering of transactions prior to any analysis. The transactions occurred during the period 1988-2000 and were for a variety of companies and consortia active in the North Sea oil and gas sector. As the majority of transactions related to a given share of the total value of an asset (a field or group of fields), the SPP represents only a proportion of the total value of the assets in question. The market values of the shares of each asset (or group of assets) involved in the 50 transactions analysed were in the range of £1m to £150m. The unadjusted total market value for these shares was approximately £2.5bn, while the unadjusted total estimated market value of all the assets involved in the 50 transactions was approximately £15bn. Each transaction involved one or more fields. Table 1 shows the year, number and combined values of the transactions in the portfolio.

**Table 1**

Date, number and Imputed Total Value (£m) of
North Sea Transactions

| 1988 | 4 | 2130 |
|------|---|------|
| 1989 | 5 | 669 |
| 1990 | 0 | 0 |
| 1991 | 4 | 1421 |
| 1992 | 5 | 1152 |
| 1993 | 5 | 1754 |
| 1994 | 6 | 2634 |
| 1995 | 8 | 1649 |
| 1996 | 2 | 441 |
| 1997 | 3 | 348 |
| 1998 | 3 | 702 |
| 1999 | 3 | 654 |
| 2000 | 2 | 1117 |

At the time of each transaction and twice during the course of this analysis, transaction data was processed through the Andersen Petroleum Services Financial Analysis Service (FAS) to produce an estimated net nominal cash flow for all fields involved in every transaction for each year of its estimated economic life. The FAS modelling system, which was based on a large (well engineered) Excel spreadsheet, used the supplied actual & projected production profiles, actual and projected oil and gas prices, estimated capex and opex, royalty, tariffs, petroleum revenue tax and corporation tax. Data was inflated using actual or estimated retail price indices. The FAS model is a deterministic DCF model and processes revenues, costs and tax etc in a straightforward manner. The net cashflows for each field involved in any given transaction were added together to produce 50 net cashflows corresponding to the 50 SPP's of each transaction.

Cashflows were discounted back to the original transaction date at either 10% for fields in production and under development or 12.5% for potentially commercial fields to yield an Estimated Market Value (EMV) for each transaction. Note that these discount rates, though arbitrary, are industry standard. Potentially commercial fields are those fields that are regarded as being likely to be developed under prevailing economic conditions within the next few years. Delays in development of these fields could be due to either technical or economic considerations. All NPV's were nominal rather than real.

Since the date of each transaction, any changes to costs, revenue, production profiles, capex, opex, exchange rates etc were recorded and the FAS model was updated to reflect these changes. The modelling software went through several revisions as the regulatory and taxation environment developed. EMV was calculated using three different data sets:

| EMV1 | At the time of the transaction | |
| --- | --- | --- |
| | Estimated Future Oil & Gas Prices | Transaction date - onwards |
| | Estimated Future Production | Transaction date - onwards |
| EMV2 | During the present Analysis | |
| | Actual Oil and Gas Prices | Transaction date - 2001 |
| | Actual Production | Transaction date - 2001 |
| | Estimated Oil & Gas Prices | 2001 onwards |
| | Estimated Future Production | 2001 onwards |
| EMV3 | During the present Analysis | |
| | Estimated Future Oil & Gas Prices | Transaction date - onwards |
| | Actual Production | Transaction date - 2001 |
| | Estimated Future Production | 2001 onwards |

EMV1 represents the best estimate of lifetime project outcome as at the time of the original transaction.

EMV2 is a more refined estimate of lifetime project outcome given current knowledge of historical oil and gas prices and production since completion of the transaction combined with up to date knowledge of the project status, regulatory and taxation regimes.

EMV3 is similar to EMV2, except that management's original estimates of oil and gas prices from transaction date to project termination were used instead of the now known values during the period transaction date-2001. The aim of this was to investigate the impact of oil price predictions available at the time of the transaction on SPP. Due to difficulties experienced in obtaining access to archived historical data, EMV3 was calculated for the 25 most recent transactions.

The oil price is Brent Crude for immediate delivery with an associated premium or discount to take into account crude quality. The gas price is based on public domain information and takes into consideration such elements as contract base year, contract base price, escalation and gas spot price.

**3. Preliminary Analysis**

As a first step, a scatter plot of SPP against EMV1 was generated. An associated linear regression was performed with SPP as the dependent variable and EMV1 as the independent variable. It is reasonable to assume that there is a causal relationship between estimated market value at the time of the transaction and the actual price paid. Figure 1 shows that market participants appeared to have paid a premium of approximately 10% to acquire the various assets. The adjusted R-squared is quite strong at 80% indicating that there was a close correlation between the NPV of the future cash flows and the sterling price paid for the assets, given the nominal discount rate used of 10% for assets in production or under development and 12.5% for potentially commercial assets. Figure 1 illustrates the nature of the relationship.

**Figure 1**

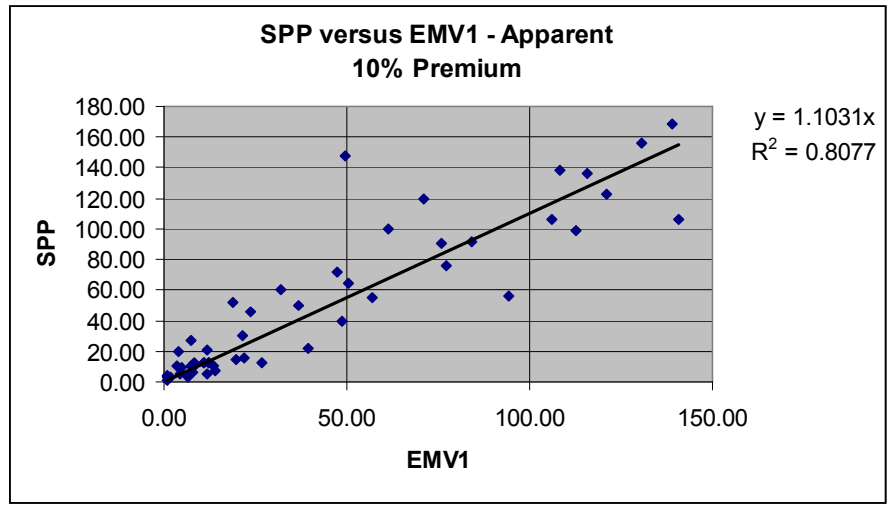

Over the course of the following years, oil and gas prices became known as did production, capex and opex, together with other information pertaining to the financial outcome of each project. EMV2, calculated in July 2001 captures this information, and augments it with best estimates regarding future happenings. A scatter plot of SPP versus EMV2 was produced, together with an associated linear regression. Again, it is reasonable to assume, given the long-term nature of oil and gas investments that there is a causative relationship between SPP and future outcome as expressed by EMV2. Not all knowledge about project outcome can be incorporated into an initial financial model. Figure 2 suggests that the market appears to have paid a 29% premium to acquire the assets given the present substantive knowledge of eventual outcome and the discount rates used. Figure 2 illustrates this.

**Figure 2**

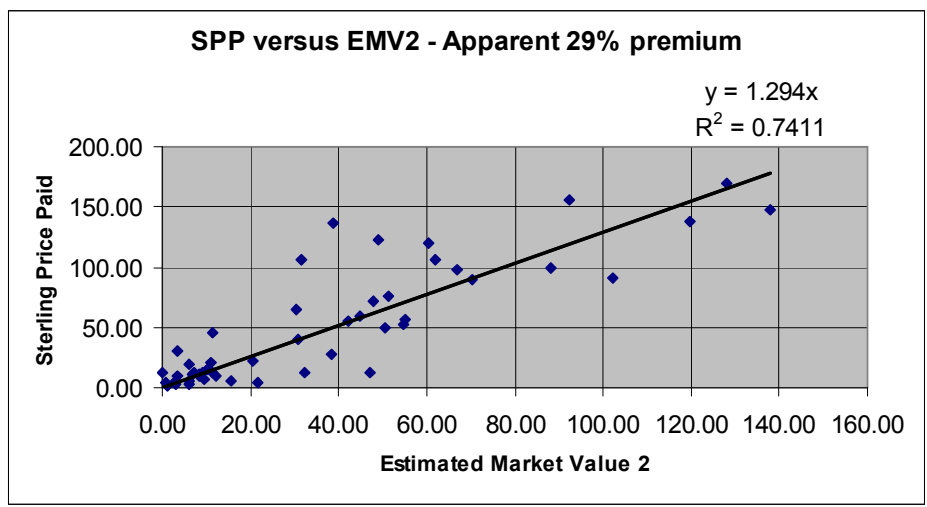

The correlation between SPP and EMV2 is not as strong as in the previous example, however it should be noted that a period of up to 12 years has elapsed between the transaction date when SPP became known and the date of calculation of EMV2. Note that in both of the prior examples several alternative regression models were investigated to see if other variables were significant including the date of the transaction, transaction size, categorised transaction size and the oil price at the time of the transaction. No other significant variables were found. In both cases the best regression was obtained by eliminating the constant and forcing the regression through the origin, as the constants of the regressions were not significant.

In the regression corresponding to Figure 2, an improvement to the fit was obtained by ignoring the three upper outliers (which reduced the premium paid to 22%). No causal basis could be established to support their removal, and a decision was made for them to remain.

Finally, since it is generally well understood that current oil prices are reflected in the investment environment, SPP was plotted against EMV3 as in Figure 3. EMV3 is identical to EMV2 except that management's' oil & gas price estimates at the time of the transaction were used. Note the extremely close correlation (80%) between SPP and EMV3 despite the smaller number of transactions (25). Removing the two upper outliers strengthens the relationship further but again, no basis could be found to support their removal. Figure 3 suggests that SPP is influenced by oil price expectations.

**Figure 3**

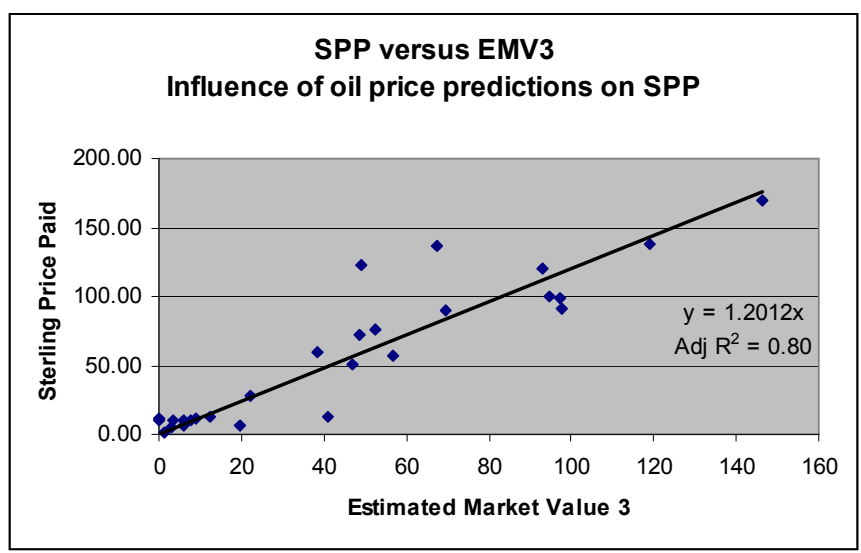

**4 The Discount Rate & Rates of Return**

One reason why SPP is systematically higher than EMV2 is that the discount rate used in the calculation of EMV2 is too high. Discount rates of 10% for fields in production or under development and 12.5% for potentially commercial fields are typically used within the industry as a means of comparing and evaluating investment opportunities. In this instance however, the actual investments and actual and future returns are known with a high degree of reliability despite the length of time elapsed between transaction date and the time of the present analysis. A rough calculation suggested that if a discount rate of approximately 6-7% had been used for the whole portfolio, SPP would on average be matched by EMV2.

In order to confirm this, the nominal annual cash flows for each transaction (corresponding to EMV2) were extracted from the FAS modelling system so that they could be studied in more detail.

Given the availability of the SPP and the subsequent actual and estimated annual cashflows of each transaction, calculating the estimated rates of return was straightforward. Figure 4 shows a frequency distribution of transaction rates of return superimposed upon the frequency distribution of the average annual Federal Reserve overnight rate from the years 1955-2001 inclusive [Federal Reserve, 2001]. The Federal Reserve overnight rate is a useful proxy for the unknown risk-free rate. Figure 5 shows a cumulative probability distribution of the two sets of returns.

As shown in Figure 4, the estimated rates of return for the 50 North Sea transactions included in this study are effectively coincident with the US risk free rate. The median rate of return for the transactions is 6.4% and is only 1% higher than the median US Federal reserve overnight rate. This is an unexpected and surprising result. The variance of the transaction returns is clearly greater than the variance of the risk free rate, however this is less surprising. Note that Federal Reserve rates for longer deposit periods are higher than the overnight rate. The oil price is denominated in US dollars.

**Figure 4**

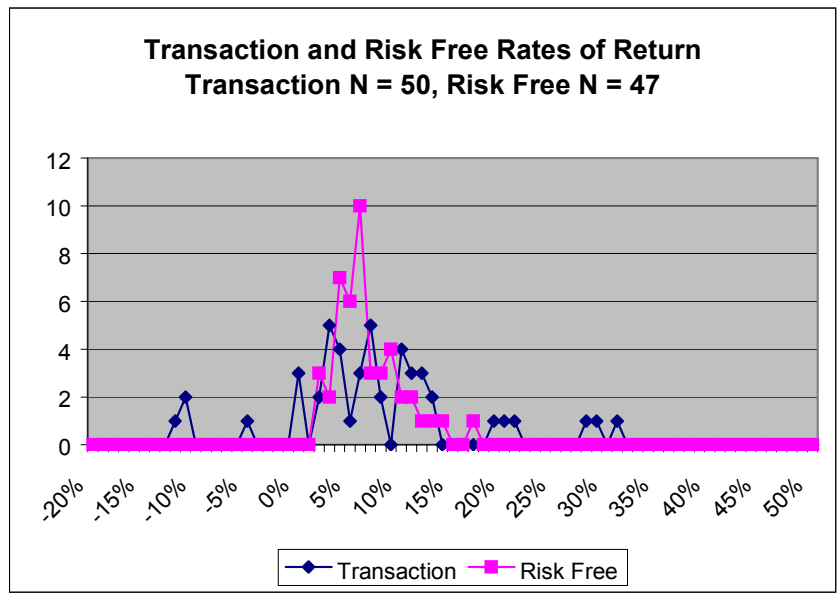

**Figure 5**

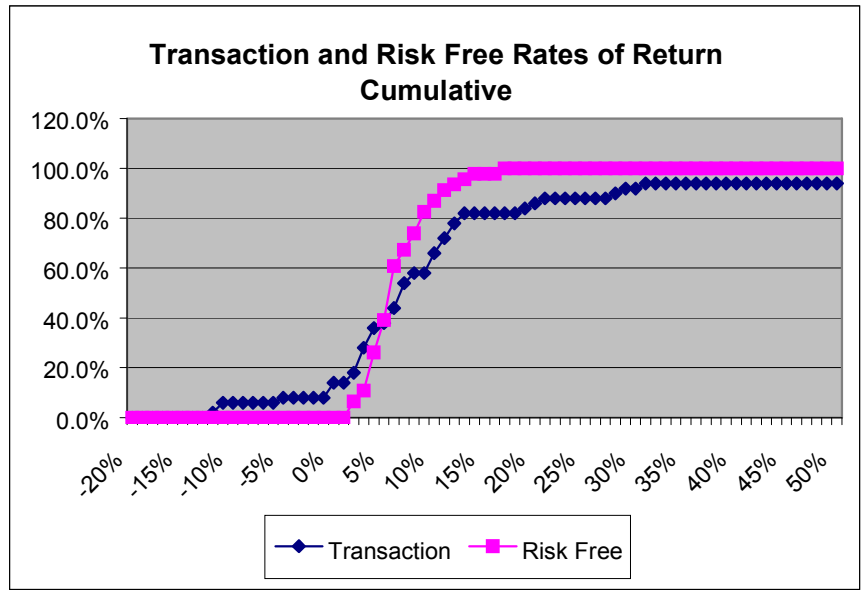

## 5 Detailed Analysis

### 5.1 Relationship between SPP & EMV1, 2 & 3

The regressions performed in section 4 may be summarised thus:

$$SPP = 1.10 * EMV1 \qquad (1)$$

$$SPP = 1.29 * EMV2 \qquad (2)$$

$$SPP = 1.20 * EMV3 \qquad (3)$$

Taking due note of the statistically estimated nature of each coefficient.

Combining equations 1 and 2, we can write:

$$K1 = \frac{EMV1}{EMV2} = \frac{1.29}{1.10} = 1.17 \qquad (4)$$

Combining equations 1 and 3 we can write:

$$K2 = \frac{EMV1}{EMV3} = \frac{1.20}{1.10} = 1.09 \qquad (5)$$

and

$$K3 = \frac{K1}{K2} = \frac{1.17}{1.09} = 1.075 \qquad (6)$$

In summary, K1 is a measure of average financial modelling performance based upon an historical analysis of a portfolio of investment opportunities. It expresses the difference between expected and actual, insofar as this can be determined, over an investment period of many years and includes the impact of oil & gas price variation. The analysis is a traditional spreadsheet based DCF using the arbitrary but industry standard discount rates of 10% and 12.5%.

K2 is similar to K1, except that the measure is independent of oil & gas price variation. K3 is an adjustment ratio that reconciles the difference between K1 and K2 and can be directly attributed to oil & gas price variation.

Regressing the 25 common data points of EMV2 and EMV3 yields a coefficient of 1.08 and an adjusted R-Squared of 0.92, supporting the above analysis.

We recalculated the internal rates of return for the 50 transactions using SPP1, which we define as:

$$SPP1 = \frac{SPP}{K3} \qquad (11)$$

Thus SPP1 represents the lower price management might have paid for the assets if management had known the oil & gas price in advance. The median rate of return for the 50 transactions was 7.8%, i.e.1.4% higher than the current estimate of the median rate of return for the transactions.

## 6 Summary

Our spreadsheet based DCF analysis of fifty historical transactions in the North Sea suggests that, at the time of the deal, the market paid, on average, an apparent initial 10% premium to acquire the assets (based on discount rates of 10% & 12.5%). However, the majority of that initial premium, 7.5% of the average asset price, was explained by over-optimism with regard to oil & gas prices.

Further analysis, based on more recent knowledge of project outcome and status suggests that the market had paid a 29% premium to acquire these assets (again based on discount rates of 10% & 12.5%). Disregarding again the estimated 7.5% effect of oil price optimism, the market still paid an apparent premium of 20%.

In fact, the apparent premia paid are entirely subjective (if not completely illusive) as the use of standard discount rates of 10% for producing fields and 12.5% for potentially commercial fields is entirely arbitrary.

We show that the median nominal rate of return for the 50 transactions in the study is 6.4%. The rate of return achieved was about 1.5% lower than what might have been achieved if oil & gas prices had followed management's initial estimates.

We also show that the median rate of return for the 50 transactions studied was 1% above a proxy for the median nominal risk-free rate.

Note that median oil price at the time of each deal in this study was $17.25. The oil price at the time of each deal varied between $11 and $29 over the 12 year period.

## 7 Conclusions

A number of tentative conclusions may be drawn from this analysis.

1. It appears to be the case that there is a strong correlation between the price the market pays for an asset and its eventual financial outcome. This would not be in any way remarkable were it not for the fact that these are capital assets being deployed in the hostile North Sea oil and gas environment over several decades. There is considerable uncertainty with regard to oil and gas prices, production, capex and opex and even exchange rates. Nevertheless, the market itself seems to be able to capture these uncertainties and reflect them consistently in the original price paid for the asset. Our analysis suggests that, at most, 25% of the price paid for an asset remains unexplained by estimated project outcome as calculated by a deterministic spreadsheet based DCF model. Spreadsheet based DCF modelling therefore seems to be quite good. Real options and other advanced decision analysis methodologies may assist in exploring the remaining 25% variation in SPP not already explained by EMV2.

2. The perceived inability of management to price projects *accurately* is inappropriate. The unexpected evidence provided in this analysis suggests that management is quite good at establishing a *consistent* price for an asset, despite the huge challenges of the North Sea environment.

3. Although management can determine a consistent price, the price management pays for an asset would appear to be rather high. On average, capital invested in the North Sea would have generated over the long term a similar but substantially more secure rate of return if it had been invested in the US Federal Reserve. Of course, the picture is incomplete because we have not studied the synergies (e.g. taxation) at company level, the optionalities from having the assets in place, informational advantages, scale economies or forecast improvements in extraction technologies and costs..

4. Management's natural inability to exactly predict the future oil price has a surprisingly small influence on the initial and eventual estimates of market value and the achieved rates of return.

5. Since the relationships studied appear to be linear, there is a suggestion that the market is indifferent to the size of a transaction. Newendorp (1975) covered the potential use of preference theory in this industry some years ago. However the evidence seems to suggest that preference theory, despite its attractions, may not be that relevant in this investment environment.

6. The stochastic nature of investment returns in the North Sea clearly supports the more widespread use of stochastic portfolio selection (Markovitz, 1952) and stochastic optimisation methods in order to spread risk and improve returns in a systematic way.

7. The common assumption that the risk free rate is fixed may be a serious limitation in certain types of investment and financial analysis, highlighting once again the Flaw of Averages (Savage, 2002).

8. Our analysis would appear to suggest that a risk neutral valuation methodology (ie free of arbitrary discount rates) would be a more appropriate basis upon which management could determine the price they wish to pay for an asset in the North Sea and elsewhere. This has the benefit of being compatible with contemporary option valuation methodologies.

9. Adoption of a risk neutral valuation methodology might resolve the issue regarding the discounting of stochastic future cashflows as discussed by Trigeorgis (1996, page 56) and Myers (1976). Given the stochastic nature of the risk free rate, stochastic actual returns and the well established notion of stochastic non-systematic or project risk (Croll, 1995) in this sector, discounting stochastic future cash flows at the Stochastic Risk Free Rate would seem to be the only way to value projects of this type (Ohlson, 1979). Stochastic Spreadsheet modelling is commonplace in this sector (Murtha, 2001).

10. Discounting at the Stochastic Risk Free Rate would in addition reduce the disadvantage long term projects tend to suffer due to compounding.

11. This work highlights some of the more recently established risks in contemporary spreadsheet modelling (Croll, 2009) including: assumptions (fixed versus stochastic risk free rate, habitual use of arbitrary discount rates); reification (ubiquity of spreadsheet based DCF modelling versus risk neutral valuation including options).

12. In a rare example of leadership by example, the spreadsheet models behind Figures 1-5 remain embedded within the word version of this paper. Oil & Gas field names, deal sizes and values, statistical information and other documentation (Payette, 2006; Pryor, 2006) are readily available for inspection. Thus the integration of a spreadsheet within its documentation have permitted the intact survival of a project over nearly a decade (Lemieux, 2005).

## Acknowledgements

We wish to express our sincere thanks to the Andersen partnership for providing the time and resources with which to complete this project and for approving the publication of our results. We thank our colleagues in the Andersen Petroleum Services and Andersen Business Modelling Groups for their assistance during the preparation of this paper.